\documentclass[journal]{IEEEtran}
\usepackage{xcolor,soul,framed} 
\colorlet{shadecolor}{yellow}
\usepackage[pdftex]{graphicx}
\graphicspath{{../pdf/}{../jpeg/}}
\DeclareGraphicsExtensions{.pdf,.jpeg,.png}
\usepackage[cmex10]{amsmath}
\usepackage{array}
\usepackage{mdwmath}
\usepackage{mdwtab}
\usepackage{eqparbox}

\usepackage{lineno}
\usepackage{amsmath}
\usepackage{graphicx}
\usepackage{float}
\usepackage{cite}
\usepackage{amssymb}
\usepackage{amsthm}
\usepackage{subfigure}
\usepackage{epsfig}
\usepackage{multirow}
\usepackage{pdfpages}
\usepackage{float}
\usepackage{nomencl}
\usepackage{subfigure}
\usepackage{epstopdf}
\usepackage{color}
\usepackage{url}
\usepackage{tabularx}
\usepackage{hhline}
\usepackage{rotating}
\usepackage{fancyhdr}
\usepackage[marginal]{footmisc}

\begin{document}
\title{DeepOPF-V: Solving AC-OPF Problems Efficiently}
\author{Wanjun~Huang,~Xiang~Pan,~Minghua~Chen,~and Steven~H.~Low
\thanks{Wanjun Huang and Minghua Chen are with School of Data Science, City University of Hong Kong. Xiang Pan is with Department of Information Engineering, The Chinese University of Hong Kong. Steven H. Low is with Department of Computing and Mathematical Sciences and Department of Electrical Engineering, California Institute of Technology. Corresponding author: Minghua Chen (Email: minghua.chen@cityu.edu.hk).}
}
\maketitle

\vspace{-1cm}
\begin{abstract}
AC optimal power flow (AC-OPF) problems need to be solved more frequently in the future to maintain stable and economic \textcolor{black}{power system operation. To tackle this challenge, a deep neural network-based voltage-constrained approach (DeepOPF-V) is proposed to solve AC-OPF problems with high computational efficiency. Its unique design predicts voltages of all buses and then uses them to reconstruct the remaining variables without solving non-linear AC power flow equations. A fast post-processing process is developed to enforce the box constraints. The effectiveness of DeepOPF-V is validated by simulations on IEEE 118/300-bus systems and a 2000-bus test system. Compared with existing studies, DeepOPF-V achieves decent computation speedup up to four orders of magnitude and comparable performance in optimality gap and preserving the feasibility of the solution.}
\end{abstract}

\begin{IEEEkeywords}
AC optimal power flow, deep neural network, voltage prediction.
\end{IEEEkeywords}
\IEEEpeerreviewmaketitle
\vspace{-0.5cm}
\section{Introduction}
\textcolor{black}{The AC optimal power flow (AC-OPF) problem is a fundamental yet challenging problem in power system operation.} With increasing uncertainties brought by intermittent renewables and highly stochastic loads, \textcolor{black}{the AC-OPF problem} needs to be solved more frequently to maintain stable and economic \textcolor{black}{power system operation. It is of great interest to solve AC-OPF problems} with high computational efficiency, especially for large-scale systems.  

Leveraging the powerful learning ability of deep neural networks (DNNs), various DNN-based approaches have been proposed to solve AC-OPF more efficiently, which can be classified into two main categories: hybrid approach and stand-alone approach. The hybrid approach aims to speed up conventional physics-based solvers (e.g., Matpower Interior Point Solver (MIPS)) by providing warm-start points to accelerate convergence\cite{dong2020smart} or predicting active \cite{chen2020learning}/inactive\cite{hasan2021hybrid} constraints to reduce problem size. Since it still needs to solve the original or truncated OPF problem iteratively, all operational constraints are considered. \textcolor{black}{However,} the speedup is limited, i.e., less than one order of magnitude in most studies. 

The stand-alone approach predicts the solution of AC-OPF directly without solving the optimization problem. Hence, it has a much greater speedup than the hybrid approach. \textcolor{black}{Following the predict-and-reconstruct framework and the handy technique of ensuring box constraints} for DC-OPF \cite{pan2019deepopf,pan2020deepopfdc}, several DNN-based approaches are developed for AC-OPF \cite{pan2020deepopfac,zamzam2020learning,baker2020emulating}. \textcolor{black}{These strategies need to solve non-linear  power flow equations in their designs,} which limits the speedup \textcolor{black}{performance} considerably. Different from the above strategies, reference \cite{chatzos2020high} combines DNNs and Lagrangian duality to predict all variables, which reported greater speedups in computation. But the critical \textcolor{black}{AC power flow equality constraints} may not be satisfied.

\textcolor{black}{In this paper, we propose DeepOPF-V as a DNN-based voltage-constrained approach to solve AC-OPF problems with high efficiency. Our contribution is threefold. First, as presented in Sec. II-B, distinct from previous studies that learn the mapping between loads and generations or all solution variables, DeepOPF-V learns the mapping between loads and voltages of all buses and directly reconstructs the remaining solution variables via simple scalar calculation, which guarantees the power flow equality constraints and is expected to achieve decent speedup\footnote{Reference \cite{rahman2020machine} also learns the mapping between loads and voltages using random forest, but it does not consider the feasibility and needs a strong assumption to enforce power balance constraints.}. Second, a fast post-processing process is developed to adjust the predicted voltages in Sec.~II-C, which helps to preserve the box constraints and improves the feasibility of the solution. Finally, we carry out simulations using IEEE 118/300-bus systems and a 2000-bus test system in Sec.~III. The empirical performance verifies the effectiveness of our design and shows that compared with existing works, DeepOPF-V achieves decent computation speedup up to four orders of magnitude and comparable performance in optimality gap and preserving the feasibility of the solution.}
\section{Model and Methodology}
\subsection{The AC Optimal Power Flow Problem}
\textcolor{black}{The} standard AP-OPF problem can be formulated as below:
\begin{align}
	&\text{min}\;\; \sum\nolimits_{i\in \mathcal{N}_G} {C(P_{gi})} \label{eq_obj} \\	
	\text{s.t.}&~P_{gi}-P_{di}=\sum\limits_{j\in\mathcal{N} }V_{i}V_{j}(g_{ij}\text{cos}\theta_{ij}+b_{ij}\text{sin}\theta_{ij}), i\in\mathcal{N}, \label{eq_Pinj} \\
	&Q_{gi}-Q_{di} =\sum_{j\in \mathcal{N}}V_{i}V_{j}(g_{ij}\text{sin}\theta_{ij}-b_{ij}\text{cos}\theta_{ij}), i\in\mathcal{N},\label{eq_Qinj} \\
	& P^{\text{min}}_{gi} \leq P_{gi} \leq P^{\text{max}}_{gi},~~i\in \mathcal{N}_G, \label{eq_pg}\\
	& Q^{\text{min}}_{gi} \leq Q_{gi} \leq Q^{\text{max}}_{gi},~~i\in \mathcal{N}_G, \label{eq_qg}\\
	& V^{\text{min}}_{i} \leq V_{i} \leq V^{\text{max}}_{i},~~i\in\mathcal{N}, \label{eq_v}\\
	& P_{ij}^{2} + Q_{ij}^{2}\leq (S^{\text{max}}_{ij})^{2},~~(i, j)\in\mathcal{E},  \label{eq_sij}\\
	& \theta^{\text{min}}_{ij} \leq \theta_{ij} \leq \theta_{ij}^{\text{max}}, ~~(i, j)\in \mathcal{E}, \label{eq_thetaij}
\end{align}
where $\mathcal{N}$, $\mathcal{N}_G$ and $\mathcal{E} $ represent the sets of all buses, generation buses and transmission lines, respectively, $g_{ij}$ and $b_{ij}$ are conductance and susceptance of branch $(i,j)$. For bus $i$, $P_{gi}$, $Q_{gi}$, $P_{di}$ and $Q_{di}$ denote active generation, reactive generation, active load and reactive load, respectively. $V_i$ and $\theta_{i}$ are voltage magnitude and angle for bus $i$, respectively, and $\theta_{ij} = \theta_i - \theta_j$. The upper and lower bounds of variable $x$ are represented by $x^{\text{max}}$ and $x^{\text{min}}$, respectively. Branch flow limit of branch $(i,j)$ is denoted as $S^{\text{max}}_{ij}$. The AC-OPF problem aims to minimize the generation costs in (\ref{eq_obj}) with all constraints satisfied. \textcolor{black}{The Kirchhoff's circuit laws are} ensured by (\ref{eq_Pinj})-(\ref{eq_Qinj}); active and reactive power generation limits are enforced by (\ref{eq_pg})-(\ref{eq_qg}); voltage magnitude limit is ensured by (\ref{eq_v}); branch flows and phase angles are restricted by (\ref{eq_sij}) and (\ref{eq_thetaij}), respectively. 
\begin{figure}[t]
	\centering
	\includegraphics[height=3.5cm, width=7cm]{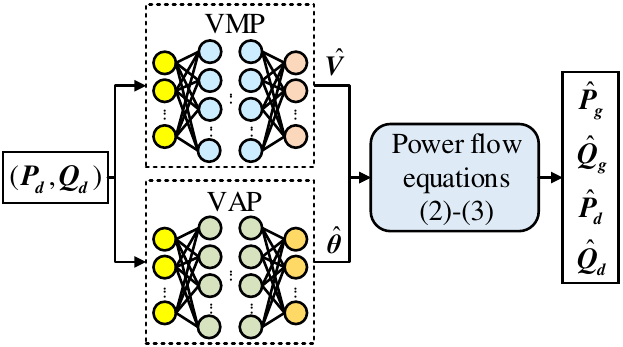}
	\caption{Schematic of the proposed DeepOPF-V.}
	\label{fig_deepopf}
\end{figure}
\vspace{-0.25cm}
\subsection{\textcolor{black}{The Proposed} DNN-Based Voltage-Constrained Approach}
Fig.~\ref{fig_deepopf} shows the schematic of the proposed DeepOPF-V. DNNs are employed to learn the mapping between loads ($\boldsymbol{P_d},\boldsymbol{Q_d}$) and voltages of all buses instead of only the generations or all the variables in previous works. {\color{black}{After training, for each input ($\boldsymbol{P_d},\boldsymbol{Q_d}$), the voltages are predicted by the well-trained DNNs instantly. \textcolor{black}{Then, using the predicted voltage magnitudes $\hat{\boldsymbol{V}}$, voltage angles $\hat{\boldsymbol{\theta}}$ and the load input ($\boldsymbol{P_d},\boldsymbol{Q_d}$), we can easily compute the right-hand side (RHS) of the equations in (\ref{eq_Pinj})-(\ref{eq_Qinj}). The remaining solution variables $\boldsymbol{\hat{P}_g}$, $\boldsymbol{\hat{Q}_g}$ and some auxiliary variables ($\boldsymbol{\hat{P}_d}$, $\boldsymbol{\hat{Q}_d}$) are then directly calculated using the obtained RHS values without the need to solve non-linear power flow equations. Specifically, for each bus $i$, 1) if there are only generators or loads, its predicted active/reactive generation (i.e., $\hat{P}_{gi}$/$\hat{Q}_{gi}$) or active/reactive load (i.e., $\hat{P}_{di}$/$\hat{Q}_{di}$) is obtained directly; 2) if there are both generators and loads, $\hat{P}_{di}$ and $\hat{Q}_{di}$ are set to the given loads $P_{di}$ and $Q_{di}$, respectively, and then $\hat{P}_{gi}$ and $\hat{Q}_{gi}$ are directly calculated from (\ref{eq_Pinj})-(\ref{eq_Qinj}). After obtaining $\boldsymbol{\hat{P}_g}$, the objective function is calculated by (\ref{eq_obj}). Due to the voltage prediction errors, there could be unsatisfied loads, i.e., the mismatches between ($\boldsymbol{P_d},\boldsymbol{Q_d}$) and ($\boldsymbol{\hat{P}_d}$, $\boldsymbol{\hat{Q}_d}$), which will be discussed in Section.~II-D.} 

To reduce the size of DNNs, $\hat{\boldsymbol{V}}$ and $\hat{\boldsymbol{\theta}}$ are predicted by voltage magnitude predictor (VMP) and voltage angle predictor (VAP), respectively, which improves the training efficiency. Hence, the inputs of VMP and VAP are both ($\boldsymbol{P_d},\boldsymbol{Q_d}$), while the outputs of VMP and VAP are $\hat{\boldsymbol{V}}$ and $\hat{\boldsymbol{\theta}}$, respectively\footnote{A lower bound for the approximation errors of the load-to-solution mapping for AC-OPF can be found in \cite{pan2020deepopfac}.}}. The loss functions of VMP (denoted as $\mathcal{L_V}$) and VAP (denoted as $\mathcal{L_\theta}$) are formulated as follows:   
\begin{align}
	\mathcal{L_V}=\sum_{i\in\mathcal{N}}||\hat{V_i}-V_i||^{2}_{2}, \quad
	\mathcal{L_\theta} =\sum_{i\in\mathcal{N}}||\hat{\theta_i}-\theta_i||^{2}_{2}, \label{eq_Loss}	
\end{align}
where $\hat{V_i}$ and $\hat{\theta_i}$ are the predicted voltage magnitude and angle for bus $i$ during the training, respectively; \textcolor{black}{$V_i$ and $\theta_i$ are the ground truths of $\hat{V_i}$ and $\hat{\theta_i}$, respectively, which can be obtained by OPF solvers such as MIPS when preparing the training data.}}

\textcolor{black}{For large power systems, we can split all buses into several groups and predict voltages for each group of buses parallelly. In this way, the DNN model size can be reduced greatly, and the training time would not increase significantly.}
\subsection{The Post-Processing Process}
{\color{black}{The post-processing process is developed to improve the feasibility of predicted solution, which contains two steps. First, the inequality constraints in (\ref{eq_pg})-(\ref{eq_thetaij}) are checked. Second, if there is violation, the related voltage magnitudes and angles will be adjusted as below:
\begin{align}
\hat{\boldsymbol{V}}_{PP}=\hat{\boldsymbol{V}} + \Delta\boldsymbol{V},~~
\hat{\boldsymbol{\theta}}_{PP} = \hat{\boldsymbol{\theta}} + \Delta\boldsymbol{\theta},\label{eq_dV}
\end{align}
where $\Delta\boldsymbol{V}$ and $\Delta\boldsymbol{\theta}$ are obtained as follows. Denote inequality constraints in (\ref{eq_pg})-(\ref{eq_thetaij}) in a compact form as $\underline{\boldsymbol{f}}\leq \boldsymbol{f}(\boldsymbol{\theta}, \boldsymbol{V})\leq \overline{\boldsymbol{f}}$. For each inequality constraint $f_{i}(\boldsymbol{\theta}, \boldsymbol{V})$, define $\Delta f_{i} = \text{max}(f_{i}(\boldsymbol{\theta}, \boldsymbol{V}) - \overline{f}_{i}, 0) + \text{min}(f_{i}(\boldsymbol{\theta}, \boldsymbol{V}) - \underline{f}_{i}, \boldsymbol{0})$. Linearizing $\Delta \boldsymbol{f}$ around the  predicted operating point ($\hat{\boldsymbol{V}}$, $\hat{\boldsymbol{\theta}}$) gives
\begin{align}
	{\left[\begin{array}{c} \Delta \boldsymbol{\theta} \\ \Delta \boldsymbol{V} \end{array}\right]} = \boldsymbol{F}_{\theta V}^{+}\Delta \boldsymbol{f},~~\boldsymbol{F}_{\theta V}={\left[\begin{array}{cc} \frac{\partial \boldsymbol{f}}{\partial\boldsymbol{\theta}} ~~ \frac{\partial\boldsymbol{f}}{\partial\boldsymbol{V}} \end{array}\right ]}, \label{eq_adjust_dV}
\end{align}
where $\boldsymbol{F}_{\theta V}^{+}$ is the pseudo-inverse of $\boldsymbol{F}_{\theta V}$. $\boldsymbol{F}_{\theta V}$ is approximated by a constant $\boldsymbol{F}_{\theta V}^{his}$ calculated at the average historical operating point $(\boldsymbol{V}^{his},\boldsymbol{\theta}^{his})$ to reduce the computational burden. \textcolor{black}{Using (\ref{eq_adjust_dV}), we adjust the values of $\Delta\boldsymbol{\theta}$ and $\Delta\boldsymbol{V}$ adaptively according to $\Delta\boldsymbol{f}$.} Note that this method helps to improve but can not guarantee the satisfaction of inequality constrains. To guarantee voltage constraints, $\hat{\boldsymbol{V}}_{PP}$ are kept within the limits after adjustment.}}
\vspace{-0.1cm}
\subsection{Load Satisfaction}
{\color{black}{Due to the prediction errors of voltages, there could be unsatisfied loads. For example, there may be mismatches between the obtained net injections and the given loads for the bus with only loads. We note that the unsatisfied loads are also inevitable in conventional approaches \cite{molzahn2014sparsity}, and 1\% load-generation imbalance is considered acceptable \cite{salgado2004reviewing}. To fully satisfy loads, controllable distributed energy sources can be applied, e.g., compensating the over/under-satisfied loads by charging/discharging batteries installed at the substation.}}
\section{Numerical Experiments}
\subsection{Experimental Setup}
Simulations are conducted on modified IEEE 118/300-bus systems \textcolor{black}{and a 2000-bus test system \cite{huang2021system}}. The dataset contains 40,000 samples for the 118-bus system and 60,000 samples for the \textcolor{black}{300/2000-bus} systems with an 80–20\% training-test split. Each sample is generated as follows. First, a load scenario is sampled randomly for each load bus from a uniform distribution of 10\% variation around the default load. Then, the loads are fed into \textcolor{black}{the conventional solver MIPS \footnote{The conventional solver MIPS solves the AC-OPF problem using the interior-point method. Other solvers can also be used to generate the training dataset.}} to obtain the optimal solutions as the ground truth values. \textcolor{black}{By learning the mapping embedded in the training dataset, DeepOPF-V can provide solutions close to those given by the solver. To verify the effectiveness of DeepOPF-V in handling systems with large correlated load variations, the real-time load data of IEEE 300-bus system in \cite{tang2017real} is used, which has load variations up to 42.3\%.}  

The DNN-based model is designed on the platform of Pytorch, which consists of fully-connected neural networks with 512,256,128-unit/1024,768,512,256-unit\textcolor{black}{/768,768,768-unit hidden layers for the 118/300/2000-bus systems}. The ReLU activation function is used on the hidden layers. We apply the Adam optimizer for training and set the maximum epoch and learning rate to 1000 and 0.001, respectively. The mini-batch size is set to 50/100/\textcolor{black}{512 for the 118/300/2000-bus systems. For the 2000-bus system, \textcolor{black}{to reduce the DNN model size, we split all buses into 10 groups evenly and predict the voltages for each group of buses parallelly.} The DNNs are trained on a single GPU, which takes 286s/1676s/692s for the 118/300/2000-bus systems. Simulation tests are} run on the quad-core (i7-3770@3.40G Hz) CPU workstation with 16GB RAM. \textcolor{black}{The codes and data are available online \cite{huang2021system}.}

The performance of DeepOPF-V is evaluated by the following metrics:
\subsubsection{\textit{Speedup}}The speedup factor $\eta_{sp}$ measures the average ratio of the computation time $t_{mips}$ consumed by MIPS to solve the original AC-OPF to the computation time $t_{dnn}$ consumed by DeepOPF-V.
\subsubsection{\textit{Optimality Loss}}It measures the average relative deviation $\eta_{opt}$ between the optimal objective value found by MIPS and that by DeepOPF-V.
\subsubsection{\textit{Constraint Satisfaction}}It evaluates the feasibility of generated solutions from two aspects: constraint satisfaction ratio (i.e., the percentage of inequality constraints satisfied) and the degree of violation (i.e., the distance between the violated variable and the boundary). The constraint satisfaction ratios (the degrees of violation) of voltage magnitude, active generation, reactive generation, branch power flow and phase angle difference are denoted by $\eta_{\boldsymbol{V}}$ ($\Delta_{\boldsymbol{V}}$), $\eta_{\boldsymbol{P_g}}$ ($\Delta_{\boldsymbol{P_g}}$), $\eta_{\boldsymbol{Q_g}}$ ($\Delta_{\boldsymbol{Q_g}}$), $\eta_{\boldsymbol{S_l}}$ ($\Delta_{\boldsymbol{S_l}}$), and $\eta_{\boldsymbol{\theta_l}}$ ($\Delta_{\boldsymbol{\theta_l}}$), respectively.
 \subsubsection{\textit{Load Satisfaction Ratio}}It is the percentage of demanded loads satisfied. The active and reactive load satisfaction ratios are denoted as $\eta_{\boldsymbol{P}_{d}}$ and $\eta_{\boldsymbol{Q}_{d}}$, respectively. 
 \begin{table}[b]
\setlength{\abovecaptionskip}{0cm}
	\centering
	\caption{Simulation Results in the 300-Bus and 2000-Bus Systems}
	\renewcommand\arraystretch{1.1}	
	\begin{tabular}{ccccc}
		\hline
		\multicolumn{1}{c}{\multirow {2}{*}{Metric}}& \multicolumn{2}{c}{IEEE 300-bus system} & \multicolumn{2}{c}{\textcolor{black}{2000-bus system}}\\
		\cline{2-5}
		\multicolumn{1}{c}{}&\multicolumn{1}{c}{Before PP} &After PP  &Before PP &After PP\\
		\hline
		$\eta_{opt}$(\%) &0.11 &0.11 &\textcolor{black}{0.15} &\textcolor{black}{0.14}  \\
		$\eta_{\boldsymbol{V}}$(\%) &100.0 &100.0 &\textcolor{black}{100.0} &\textcolor{black}{100.0} \\
		$\eta_{\boldsymbol{P_g}}$(\%)/$\eta_{\boldsymbol{Q_g}}$(\%) &99.9/99.3 &100.0/99.8 &\textcolor{black}{100.0/100.0} &\textcolor{black}{100.0/100.0}\\	
		$\Delta_{\boldsymbol{P_g}}$(p.u.) &0.0020 &0.0020 &\textcolor{black}{0} &\textcolor{black}{0}\\
		$\Delta_{\boldsymbol{Q_g}}$(p.u.) &0.3350 &0.3350 &\textcolor{black}{0} &\textcolor{black}{0}\\
		$\eta_{\boldsymbol{S_l}}$(\%) &100.0 &100.0  &\textcolor{black}{99.71} &\textcolor{black}{99.71} \\
		$\Delta_{\boldsymbol{S_l}}$(p.u.) &0 &0 &\textcolor{black}{0.0247} &\textcolor{black}{0.0247}\\
		$\eta_{\boldsymbol{\theta_l}}$(\%) &100.0 &100.0  &\textcolor{black}{100.0} &\textcolor{black}{100.0}\\
		$\eta_{\boldsymbol{P}_{d}}$(\%)/$\eta_{\boldsymbol{Q}_{d}}$(\%)  &99.6/99.5 &99.6/99.4 &\textcolor{black}{99.83/99.53} &\textcolor{black}{99.84/99.53}\\
		$t_{mips}$/$t_{dnn}$ (ms) &3213.3/1.7 &3213.3/2.1 &\textcolor{black}{39107.8/2.7} &\textcolor{black}{39107.8/2.9}\\
		$\eta_{sp}$ &$\times$1890 &$\times$1530 &\textcolor{black}{$\times$16543} &\textcolor{black}{$\times$15374}\\
		\hline  
	\end{tabular}\label{tab_case30_300}
\end{table}
\subsection{Performance Evaluation}
\subsubsection{\textcolor{black}{Simulation results in the 300/2000-bus systems}} \textcolor{black}{The results in Table.~\ref{tab_case30_300} indicate that DeepOPF-V can speed up the solution of AC-OPF significantly (i.e., up to three/four orders of magnitude in the 300/2000-bus systems) with negligible optimality loss (i.e., less than 0.15\%).} Besides, almost all loads are satisfied. As for the feasibility of the solution, voltage and phase angle constraints are all satisfied. For the 300-bus system, the active generation and branch flow constraints are all satisfied after \textcolor{black}{post-processing (PP)}. Almost all reactive generation constraints are satisfied with negligible violation degrees that only account for 0.03\% capacities of the largest generators installed in the system. \textcolor{black}{For the 2000-bus system, the generation constraints are all satisfied. Since most branch flow constraints are already binding in the dataset, they are slightly violated due to the prediction errors, within 1\% MVA rating of the branch on average.}

\textcolor{black}{Table.~\ref{tab_300_daily_data} shows that DeepOPF-V achieves high efficiency when load variations are significant (up to 42.3\%) and correlated. Besides, the performances of using $\boldsymbol{F}_{\theta V}$ and the approximated matrix $\boldsymbol{F}_{\theta V}^{his}$ for the PP are compared. As seen in Table.~\ref{tab_300_daily_data}, these two methods have the same performance except for the computational speedup. One the of main reasons is that DeepOPF-V has high prediction accuracy (i.e., the mean square prediction errors of voltage magnitudes and angles are 9.03e-5 p.u. and 2.79e-4 p.u., respectively). Thus, it still achieves good performance without PP.}

\begin{table}[t]
\setlength{\abovecaptionskip}{0.cm}
\setlength{\belowcaptionskip}{-0.5cm}
	\centering
	\caption{\textcolor{black}{Simulation Results in the Modified IEEE 300-Bus System with Real-Time Load Data}}
	\renewcommand\arraystretch{1.1}	
	\begin{tabular}{cccc}		
		\hline 
		\multicolumn{1}{c}{\multirow {2}{*}{\textcolor{black}{Metric}}} &\multicolumn{1}{c}{\multirow {2}{*}{\textcolor{black}{Before PP}}} &\multicolumn{2}{c}{\textcolor{black}{After PP}}\\
		\cline{3-4}
		\multicolumn{1}{c}{} &\multicolumn{1}{c}{} &\textcolor{black}{$\boldsymbol{F}_{\theta V}^{his}$} &\textcolor{black}{$\boldsymbol{F}_{\theta V}$}\\
		\hline 
		\textcolor{black}{$\eta_{opt}$(\%)}/\textcolor{black}{$\eta_{\boldsymbol{V}}$(\%)} &\textcolor{black}{-0.01}/\textcolor{black}{100.0} &\textcolor{black}{-0.01}/\textcolor{black}{100.0} &\textcolor{black}{-0.01}/\textcolor{black}{100.0} \\
		\textcolor{black}{$\eta_{\boldsymbol{P_g}}$(\%)/$\Delta_{\boldsymbol{P_g}}$(p.u.)} &\textcolor{black}{99.6/0.0007} &\textcolor{black}{100.0/0} &\textcolor{black}{100.0/0} \\
		\textcolor{black}{$\eta_{\boldsymbol{Q_g}}$(\%)/$\Delta_{\boldsymbol{Q_g}}$(p.u.)} &\textcolor{black}{99.8/0.0019} &\textcolor{black}{100.0/0}  &\textcolor{black}{100.0/0}\\ 
		\textcolor{black}{$\eta_{\boldsymbol{S_l}}$(\%)/$\eta_{\boldsymbol{\theta_l}}$(\%)} &\textcolor{black}{100.0/100.0} &\textcolor{black}{100.0/100.0} &\textcolor{black}{100.0/100.0}\\
		\textcolor{black}{$\eta_{\boldsymbol{P_d}}$(\%)/$\eta_{\boldsymbol{Q_d}}$(\%)} &\textcolor{black}{99.90/99.90} &\textcolor{black}{99.95/99.94} &\textcolor{black}{99.95/99.94}\\
		\textcolor{black}{$\eta_{sp}$} &\textcolor{black}{$\times$1887}   &\textcolor{black}{$\times$1562} &\textcolor{black}{$\times$647}\\
		\hline  			
	\end{tabular} \label{tab_300_daily_data}
\end{table}
\subsubsection{Comparison with state-of-the-art approaches} The proposed DeepOPF-V is also compared with the state-of-the-art approaches in \cite{pan2020deepopfac} (denoted as DACOPF) and \cite{baker2020emulating} (denoted as EACOPF). The parameters of DACOPF and EACOPF are set according to \cite{pan2020deepopfac} and \cite{baker2020emulating}, respectively. For a fair comparison, \textcolor{black}{the training/testing samples are all set to 32,000/8,000. These approaches are not compared in larger systems for the excessive training time of the solutions in \cite{pan2020deepopfac,baker2020emulating}.}

As shown in Table.~\ref{tab_compare_case118}, all approaches have small optimality losses. However, DeepOPF-V has a much larger speedup (around three orders of magnitude) than the other approaches (around one order of magnitude). As for the feasibility of the solution, there is no violation of inequality constraints in DeepOPF-V, while there are voltage magnitude constraints violated in DACOPF and voltage and active generation constraints violated in EACOPF. That is because DeepOPF-V obtains voltages directly and thus can keep them within limits, while DACOPF and EACOPF obtain voltages by solving power flow equations using predicted generation set points and therefore can not ensure voltage constraints. Moreover, there may be no feasible power flow solutions in DACOPF and EACOPF, {\color{black}{which is not a concern in DeepOPF-V because power flow equations are satisfied automatically}}.
\begin{table}[t]
\setlength{\abovecaptionskip}{0.cm}
\setlength{\belowcaptionskip}{-0.5cm}
	\centering
	\caption{Comparison Results in Modified IEEE 118-Bus System}
	\renewcommand\arraystretch{1.1}	
	\begin{tabular}{cccc}		
		\hline 
		Metric  &DeepOPF-V  &DACOPF &EACOPF \\	
		\hline 
		$\eta_{opt}$(\%) &0.1 &0.5  &0.4\\ 
		$\eta_{\boldsymbol{V}}$(\%) /$\Delta_{\boldsymbol{V}}$(p.u.) &100.0/0 &98.6/0.0074  &99.2/0.0043\\ 
		$\eta_{\boldsymbol{P_g}}$(\%)/$\Delta_{\boldsymbol{P_g}}$(p.u.) &100.0/0 &100.0/0  &99.3/0.0155\\
		$\eta_{\boldsymbol{Q_g}}$(\%) &100.0 &100.0  &100.0\\ 
		$\eta_{\boldsymbol{S_l}}$(\%)/$\eta_{\boldsymbol{\theta_l}}$(\%) &100.0/100.0 &100.0/100.0 &100.0/100.0\\
		$\eta_{\boldsymbol{P_d}}$(\%)/$\eta_{\boldsymbol{Q_d}}$(\%) &99.8/99.6 &100.0/100.0 &100.0/100.0\\
		$\eta_{sp}$ &around $\times$1000   &around $\times$10 &around $\times$10 \\
		\hline  			
	\end{tabular} \label{tab_compare_case118}
\end{table}

All loads can be fully satisfied in DACOPF and EACOPF so long as there are power flow solutions. However, there is no guarantee of the existence of power flow solutions. In this test system, 0.15\% testing samples have no power flow solutions in EACOPF. In contrast, DeepOPF-V always guarantees to obtain power flow solutions, with a load satisfaction ratio of 99.6\%, {\color{black}{which is acceptable as discussed in Section II.D.}}

\section{Conclusion}
\textcolor{black}{We propose} a DNN-based voltage-constrained approach (DeepOPF-V) to solve AC-OPF problems with high computational efficiency. \textcolor{black}{It} predicts voltages of all buses using a DNN-based model and then obtains all remaining variables via power flow equations, which ensures voltage and power balance constraints. Simulation results on \textcolor{black}{IEEE 118/300-bus systems and a 2000-bus test system indicate that DeepOPF-V outperforms the state-of-the-art approaches in computation speedup (up to four orders of magnitude faster than conventional solver), with similar performance in optimality loss (less than 0.2\%) and preserving the feasibility of the solution.} 

\ifCLASSOPTIONcaptionsoff
  \newpage
\fi
\bibliographystyle{IEEEtran}
\bibliography{IEEEabrv,Bibliography}
\vfill
\end{document}